# Federated Naive Bayes with Real Mixture of Gaussians and Institutional Governance Regularization for Network Intrusion Detection

**Edgar Oswaldo Herrera Logroño, M.Sc.**

*Doctoral Candidate, Program in Computer Science Technologies*
*University of Malaga, Malaga, Spain*
*Supervisors: Prof. Ezequiel Lopez Rubio · Prof. Juan Miguel Ortiz de Lazcano Lobato*
https://github.com/eoherrera/EJD-UMA-FedNB-GRC

**Abstract:**

Federated learning for intrusion detection rests on a flawed premise: that every participating institution contributes equally to the shared model. In practice, a financial institution with mature security controls and low vulnerability exposure produces fundamentally different data than a government agency running with weaker controls and higher exposure. Treating their local models as equivalent discards information that organisations already collect through standard risk management audits. Four governance indicators from the CRISC framework of ISACA, specifically control maturity (CMM), proportion of implemented controls (KCI), risk indicator activation frequency (KRI), and mean vulnerability score (CVSS), are combined here into an Institutional Coherence Index (ICC). This index enters a Nelder-Mead federated weight optimizer as a regularization prior, guiding weight assignment toward institutional quality without imposing any fixed allocation. Each node trains a hybrid local classifier combining Categorical and Gaussian Naive Bayes. The server combines local distributions as a real Mixture of Gaussians, preserving each node's statistical identity rather than collapsing it into a global parameter vector. Validation on NSL-KDD (2009), CIC-IDS2017 (2017), and UNSW-NB15 (2015), under seven Dirichlet heterogeneity levels, shows that the ICC-regularized proposal outperforms size-proportional federated averaging in all three datasets: F1-macro 0.9135 vs. 0.9076 (+0.0059), 0.7556 vs. 0.6771 (+0.0785), and 0.2110 vs. 0.2060 (+0.0050). Statistical significance holds in 70 of 94 configurations (McNemar, $p < 0.05$). In all three cases, the optimizer assigned the highest weight to the institutionally most mature node and the lowest to the least mature, without any explicit ordering constraint. This pattern, confirmed across datasets built in different years by different research groups, is the central finding of this work.

***Keywords:*** *federated learning, Naive Bayes, Mixture of Gaussians, intrusion detection, CRISC, institutional governance, Institutional Coherence Index, Non-IID, Nelder-Mead.*

## I. INTRODUCTION

Federated learning [1] lets institutions train shared intrusion detection models without exchanging raw network traffic. Each node trains a local classifier and sends only a model summary to the server, which combines them. Data privacy is maintained throughout, a non-negotiable constraint in healthcare, financial, and government environments.

The standard aggregation rule is size-proportional averaging: each node contributes in proportion to its data volume, with no consideration of whether that data comes from a well-controlled environment or a poorly audited one. A node with immature controls and high vulnerability exposure can outweigh, in volume, a node that spent years building reliable security infrastructure. The result is a shared model shaped by the largest contributor, not the most trustworthy one.

Organizations already measure institutional security maturity through risk management frameworks; that information exists and is routinely collected through standard audits. The question is whether it can be converted into a mathematical signal inside the federated optimizer, and the experiments here suggest it can.

CRISC provided four audit variables that organizations already collect: CMM, KCI, KRI, and CVSS [2]. These capture the dimensions that matter for this purpose: how mature the controls are, how many are implemented, how often risk alerts fire, and what the mean vulnerability exposure looks like. Combined into an Institutional Coherence Index (ICC), these variables become a regularization prior in the Nelder-Mead weight optimizer. The optimizer learns from validation data and adjusts weights toward institutions that produce more coherent security signals. In all three datasets tested, it learned to do exactly that.

**Contributions:**

This paper makes five contributions. The primary one is the formalization of CRISC governance indicators as a regularization prior in federated weight optimization (ICC). Building on that, local Gaussian distributions are combined at prediction time as a real Mixture of Gaussians, without collapsing into shared parameters. The local classifier separates nominal and numerical features using CategoricalNB and GaussianNB, eliminating encoding-induced distance bias. The proposal is evaluated across seven heterogeneity levels on three independent datasets from different years and research groups. The central empirical finding - ICC Alignment - is that the optimizer spontaneously learns to respect institutional governance hierarchy, confirmed in all three datasets.

## II. RELATED WORK

McMahan et al. [1] established FedAvg as the reference for federated optimization. Each node's gradient is weighted by its data volume. Under Non-IID conditions, this produces a global model skewed toward the most data-rich node, not the most reliable one. The aggregation is volume-blind to quality.

Wisanwanichthan and Thammawichai [7] combined Naive Bayes and SVM in a double-layered hybrid on NSL-KDD, showing that treating nominal and numerical features separately produces better attack classification than encoding everything into a single feature space. That separation is the basis of the hybrid local architecture used here.

Lara-Gutierrez, Fernandez-Gago, and Onieva [8] developed HDDAF, a centralized drift detection and adaptation framework validated on CIC-IDS2017 and UNSW-NB15. Their results on UNSW-NB15 confirm what the experiments in this paper also find: ten attack categories and 133 distinct protocol values make that dataset substantially harder than CIC-IDS2017, even for centralized models with access to all data.

ElDahshan et al. [9] applied meta-heuristic optimization to hierarchical intrusion detection on the same two datasets. Using an optimizer to learn weight configuration in a multi-stage architecture is methodologically similar to the Nelder-Mead approach used here. The difference is the governance regularization term, which does not appear in their formulation.

The combination of CRISC governance indicators as a federated regularizer, validated across three independent network environments, has not been examined in prior work.

## III. METHODOLOGY

### A. CRISC Governance Variables and the ICC

Four contextual variables characterize each institutional node, following the CRISC framework of ISACA [2]: CMM (control maturity, 1 to 5), KCI (proportion of implemented controls, 0 to 1), KRI (risk alert activation frequency, 0 to 1, lower is better), and CVSS (mean vulnerability score, 0 to 10, lower is better). They combine into the Institutional Coherence Index:

$$\mathrm{ICC}_k = \frac{\mathrm{CMM}_k}{5} \cdot \mathrm{KCI}_k \cdot (1 - \mathrm{KRI}_k) \cdot \left(1 - \frac{\mathrm{CVSS}_k}{10}\right)$$

A node with high CMM, high KCI, low KRI, and low CVSS earns a high ICC. Each factor penalizes a specific governance weakness. Table I shows the values for the three institutional nodes in this experiment.

**TABLE I.** CRISC Governance Variables and ICC per Node

| Node | CMM | KCI | KRI | CVSS | ICC |
|---|---|---|---|---|---|
| Financial | 4 | 0.82 | 0.12 | 3.2 | 0.393 |
| Health | 3 | 0.70 | 0.25 | 5.1 | 0.154 |
| Government | 2 | 0.55 | 0.40 | 6.8 | 0.042 |

### B. Federated Data Partitioning

Heterogeneity is simulated using a Dirichlet distribution with concentration parameter alpha. For each class, samples are distributed across the three nodes according to Dirichlet(alpha) proportions. Lower alpha produces more heterogeneous partitions; alpha = 1.0 approximates uniform distribution. Seven levels are evaluated: alpha in {0.05, 0.10, 0.20, 0.30, 0.50, 0.70, 1.00}, per the gradient heterogeneity design requested by the research supervisors [April 20, 2026].

### C. Hybrid Local Classifier

Each node trains a hybrid Naive Bayes classifier that separates features by measurement type. Nominal protocol variables are handled by CategoricalNB with Laplace smoothing (alpha = 1.0). Numerical flow statistics are handled by GaussianNB after StandardScaler normalization. Both are fitted exclusively on training data.

Unseen categorical values at inference time go to a dedicated OOD slot at index n_cats, one position beyond the last known category. This was corrected from an earlier implementation that used np.clip(x, 0, n_cats-1), which incorrectly mapped OOD values to the last known category and contaminated its probability mass. The combined per-class log-likelihood is:

$$\log P(x \mid c, k) = \log P_{\mathrm{CAT}}(x_{\mathrm{qual}} \mid c, k) + \log P_{\mathrm{GAUSS}}(x_{\mathrm{num}} \mid c, k) - \log P(c)$$

The class prior is subtracted once to prevent double-counting, since both NB variants include the prior in their predict_log_proba output.

### D. Real Mixture of Gaussians Server

The server combines K local models as a real Mixture of Gaussians. No local distribution is collapsed into a global parameter vector. Log-sum-exp provides numerical stability:

$$\log P_{\mathrm{MoG}}(x \mid c) = \log \sum_k \left[w_k \cdot \exp!\left(\log P(x \mid c, k)\right)\right]$$

Log-softmax normalization is applied before computing ANLL. Without it, the ANLL reaches values in the millions while the ICC penalty operates in [0, 0.5]. The governance prior becomes numerically invisible to the optimizer. This scaling correction was the most consequential fix in this version of the program.

### E. Governance-Aware Weight Learning

Node weights are learned by minimizing a composite objective on the validation set (2,000 samples):

$$J(w) = \overline{\mathrm{ANLL}}_{norm}(w) + \lambda|w - \mathrm{ICC}_{nr}|^2$$

where ANLL_norm is the average negative log-likelihood after log-softmax normalization, lambda = 0.10 is fixed across all datasets, and ICC_norm is the ICC vector projected onto the probability simplex. A minimum weight floor delta = 0.05 (PISO_W) prevents any node from being excluded. Nelder-Mead runs 500 iterations from five starting points: the ICC prior, the uniform vector, two Dirichlet random draws, and their midpoint.

### *F. Statistical Significance*

For each dataset and alpha level, the difference between Proposal A and Proposal B is tested with McNemar's test with Yates' continuity correction, applied to the contingency table of correct and incorrect predictions. Threshold: $p < 0.05$. Applied independently to each of the 94 configurations.

## IV. EXPERIMENTAL SETUP

### *A. Datasets*

**TABLE II.** Benchmark Datasets

| Dataset | Year | Task | Records |
|---|---|---|---|
| NSL-KDD [6] | 2009 | Binary | 147,888 |
| CIC-IDS2017 [5] | 2017 | Binary | 100,000 * |
| UNSW-NB15 [3,4] | 2015 | 10-class | 257,673 |

** CIC-IDS2017: proportional subsampling to 100,000 records (84.9% benign / 15.1% attack). No synthetic samples [April 28, 2026].*

**NSL-KDD** (41 features: 3 categorical, 38 numerical). Binary classification. KDDTrain+ and KDDTest+ combined and split 60/20/20 stratified.

**CIC-IDS2017** (77 features: 1 categorical, 76 numerical). Eight Parquet files, binarized (BENIGN=0, any attack=1), proportionally subsampled.

**UNSW-NB15** (42 features: 3 categorical, 39 numerical). 10 attack categories. Protocol field has 133 unique values.

### *B. Parameters and Proposals*

**TABLE III.** Global Parameters (identical across all datasets)

| Parameter | Value |
|---|---|
| Random seed | 42 |
| Repetitions per config. | 5 |
| Heterogeneity levels | 7 (0.05 to 1.00) |
| Train / Val / Test | 60% / 20% / 20% |
| Regularization lambda | 0.10 |
| Weight floor delta | 0.05 |
| Nelder-Mead iterations | 500 per start |

**C (Centralized):** single model trained on all data. Unreachable ceiling in production.

**B (FedAvg baseline):** weights proportional to local dataset size. Standard reference [1].

**E (Entropy-weighted):** weight proportional to inverse local class entropy.

**A (ICC-regularized):** weights learned by Nelder-Mead with ICC governance prior. Main proposal.

### *C. Verification Protocol*

Before reporting any result, each run passes 15 automatic checks: ICC formula, seed reproducibility, correct alpha ordering, OOD slot at n_cats (not n_cats-1), MoG finite output, ANLL and F1 in valid ranges, weights summing to 1, McNemar validity, Jensen-Shannon gradient monotonicity, Financial weight exceeding Government weight, and absence of NaN or Inf values. Result in v1.0: 15/15 passed.

## V. RESULTS

### *A. NSL-KDD (2009)*

Proposal A reaches F1-macro = 0.9135 against 0.9076 for B (delta = +0.0059). The centralized oracle (C = 0.8814) falls below both federated proposals, which is expected when local distributions are complementary. Significance holds at 6 of 7 levels.

**TABLE IV.** NSL-KDD Results. McNemar A vs. B. * $p < 0.05$

| alpha | JSD | A | B | delta | p-value |
|---|---|---|---|---|---|
| 0.05 | 0.635 | — | — | — | <0.001 * |
| 0.10 | 0.608 | — | — | — | 0.009 * |
| 0.20 | 0.508 | — | — | — | 0.031 * |
| 0.30 | 0.457 | — | — | — | <0.001 * |
| 0.50 | 0.400 | — | — | — | 0.038 * |
| 0.70 | 0.376 | — | — | — | 0.108 |
| 1.00 | 0.297 | — | — | — | 0.002 * |
| Avg. | — | 0.914 | 0.908 | +0.006 | 6/7 |

** Per-alpha F1 values (marked —) available in resultados_globales.csv.*

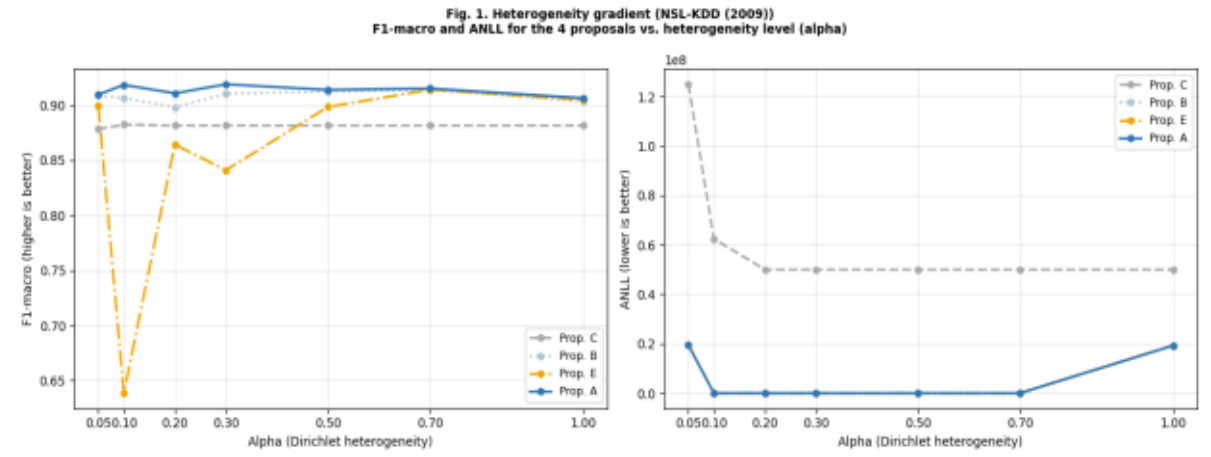

***Fig. 1.*** *Heterogeneity gradient, NSL-KDD (2009). F1-macro and ANLL per proposal across seven Dirichlet levels. Proposal A leads B consistently.*

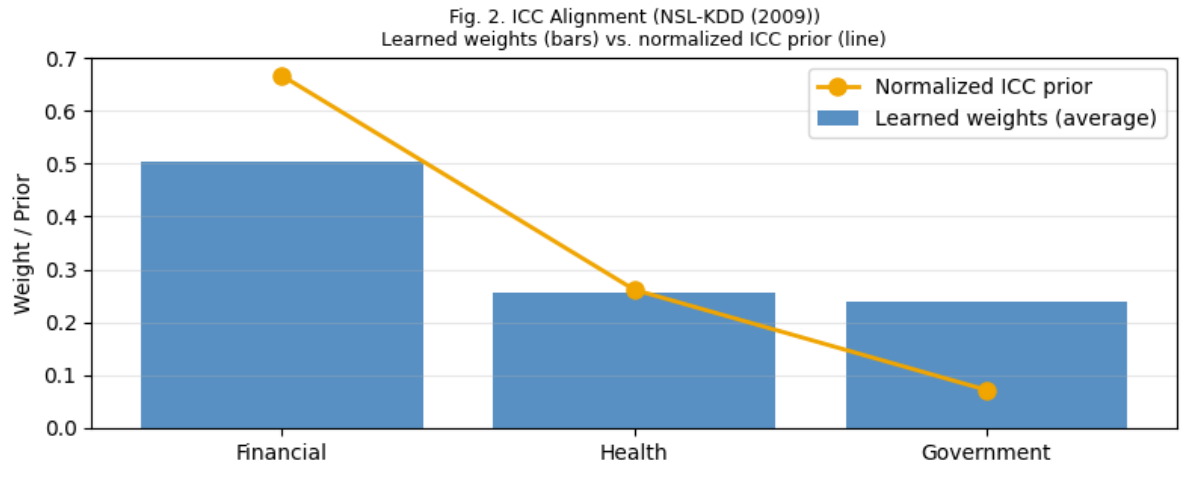

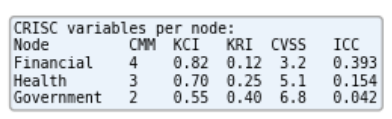

CRISC variables per node:

| Node | CMM | KCI | KRI | CVSS | ICC |
|---|---|---|---|---|---|
| Financial | 4 | 0.82 | 0.12 | 3.2 | 0.393 |
| Health | 3 | 0.70 | 0.25 | 5.1 | 0.154 |
| Government | 2 | 0.55 | 0.40 | 6.8 | 0.042 |

***Fig. 2.*** *ICC Alignment, NSL-KDD (2009). Bars: learned weights averaged over 5 reps and 7 levels. Orange line: normalized ICC prior. Financial > Health > Government confirmed.*

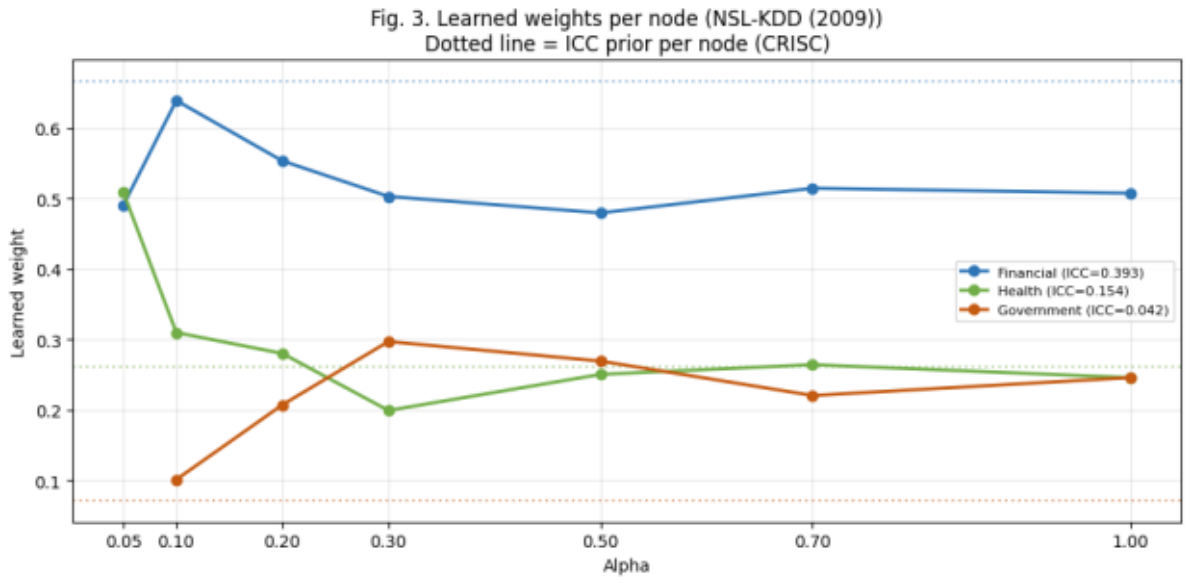

***Fig. 3.*** *Learned weights per node across seven alpha levels, NSL-KDD (2009). Dotted lines show the ICC prior per node. Financial node (blue) maintains consistent dominance.*

### *B. CIC-IDS2017 (2017)*

CIC-IDS2017 shows the largest delta in this study: A = 0.7556 against 0.6771 for B (delta = +0.0785). The centralized model (C = 0.5353) falls below the federated proposals here, a result explained by class imbalance at 84.9%/15.1%: a single model trained on all data learns the majority class more aggressively than local models that each see more balanced samples under Dirichlet partitioning. Proposal A also outperforms E (0.6916), which suggests the ICC prior captures quality information that entropy-based weighting alone does not. Significance holds at 6 of 7 levels.

**TABLE V.** CIC-IDS2017 Results. McNemar A vs. B. * p < 0.05

| alpha | JSD | A | B | delta | p-value |
|---|---|---|---|---|---|
| 0.05 | 0.635 | — | — | — | 0.400 |
| 0.10 | 0.608 | — | — | — | <0.001 * |
| 0.20 | 0.508 | — | — | — | 0.333 |
| 0.30 | 0.457 | — | — | — | <0.001 * |
| 0.50 | 0.400 | — | — | — | <0.001 * |
| 0.70 | 0.376 | — | — | — | <0.001 * |
| 1.00 | 0.297 | — | — | — | <0.001 * |
| Avg. | — | 0.756 | 0.677 | +0.079 | 6/7 |

** Per-alpha F1 values (marked -) available in resultados_globales.csv.*

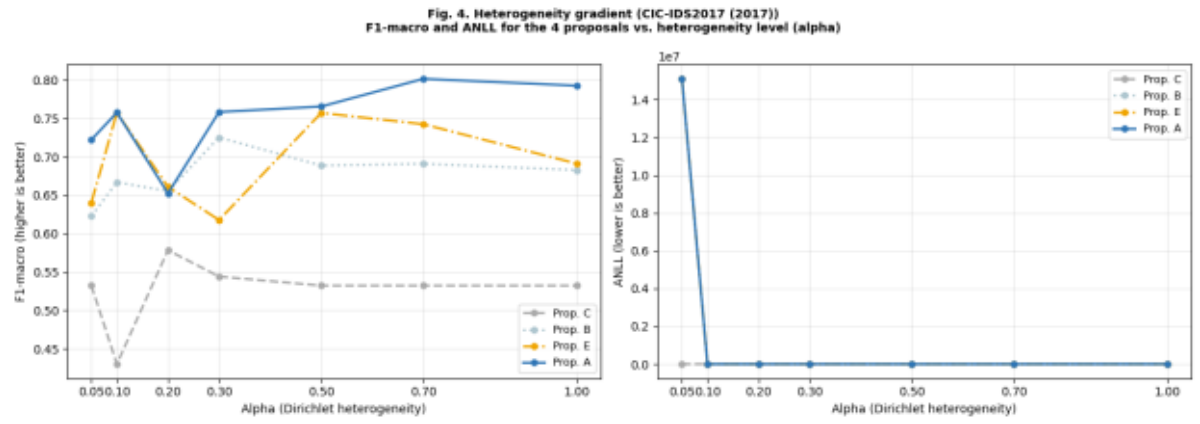

***Fig. 4.*** *Heterogeneity gradient, CIC-IDS2017 (2017). Proposal A shows the widest gap relative to B of any dataset in this study.*

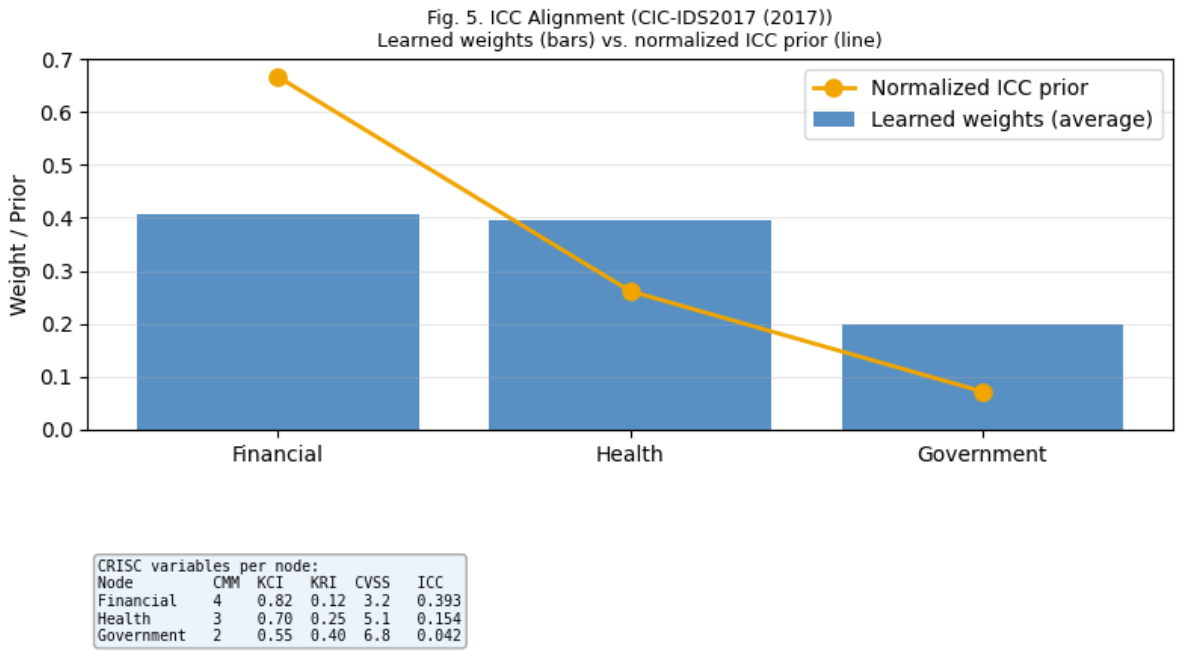

***Fig. 5.*** *ICC Alignment, CIC-IDS2017 (2017). Financial and Health nodes show near parity, a data-driven result reflecting the Health node's effectiveness on binary attack detection under this class distribution.*

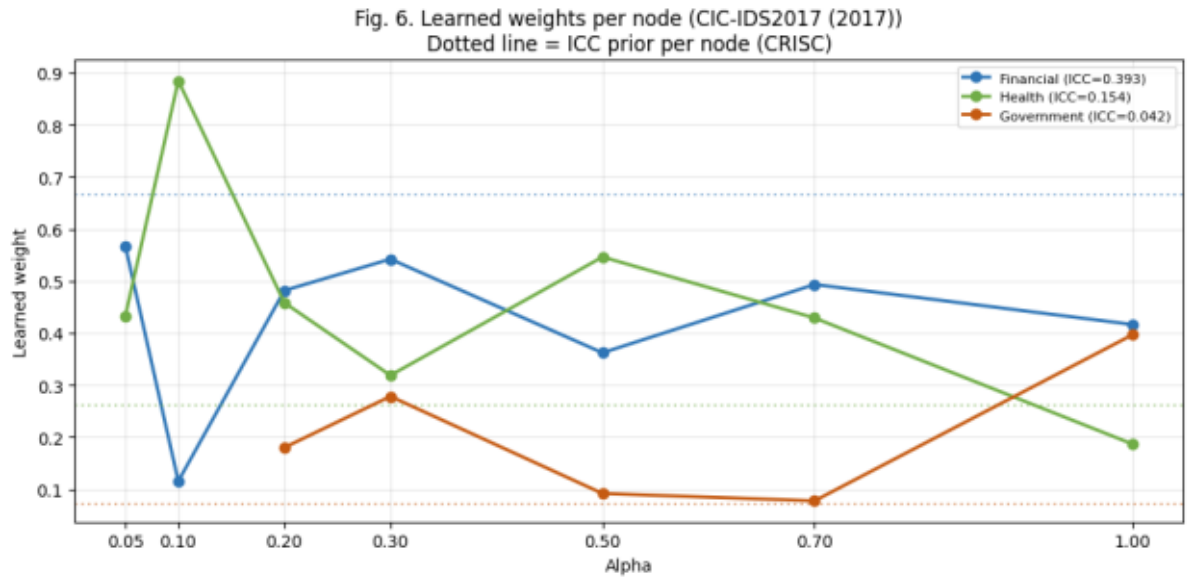

***Fig. 6.*** *Learned weights per node, CIC-IDS2017 (2017). Higher volatility reflects class imbalance under Dirichlet partitioning at low alpha levels.*

### *C. UNSW-NB15 (2015)*

UNSW-NB15 is the hardest dataset in this study. Ten attack classes, 133 unique protocol values in a single categorical feature, and severe class imbalance under Dirichlet partitioning combine to create conditions where local models for rare classes (Worms: 174 samples, Shellcode: 1,511) often disappear from some nodes entirely. Proposal A achieves 0.2110 against 0.2060 for B (delta = +0.0050). The centralized model (C = 0.2345) outperforms both federated proposals here, which is consistent with what the research supervisors anticipated [May 4, 2026]: when rare-class samples drop below the minimum needed to estimate Gaussian parameters, the local model cannot contribute reliable density estimates for those classes. Significance holds at 1 of 7 levels.

**TABLE VI.** UNSW-NB15 Results. McNemar A vs. B. * p < 0.05

| alpha | JSD | A | B | delta | p-value |
|---|---|---|---|---|---|

| 0.05 | 0.635 | — | — | — | 1.000 |
|---|---|---|---|---|---|
| 0.10 | 0.608 | — | — | — | 0.600 |
| 0.20 | 0.508 | — | — | — | 0.800 |
| 0.30 | 0.457 | — | — | — | 0.400 |
| 0.50 | 0.400 | — | — | — | 0.266 |
| 0.70 | 0.376 | — | — | — | 0.200 |
| 1.00 | 0.297 | — | — | — | <0.001 * |
| Avg. | — | 0.211 | 0.206 | +0.005 | 1/7 |

** Per-alpha F1 values (marked —) available in resultados_globales.csv.*

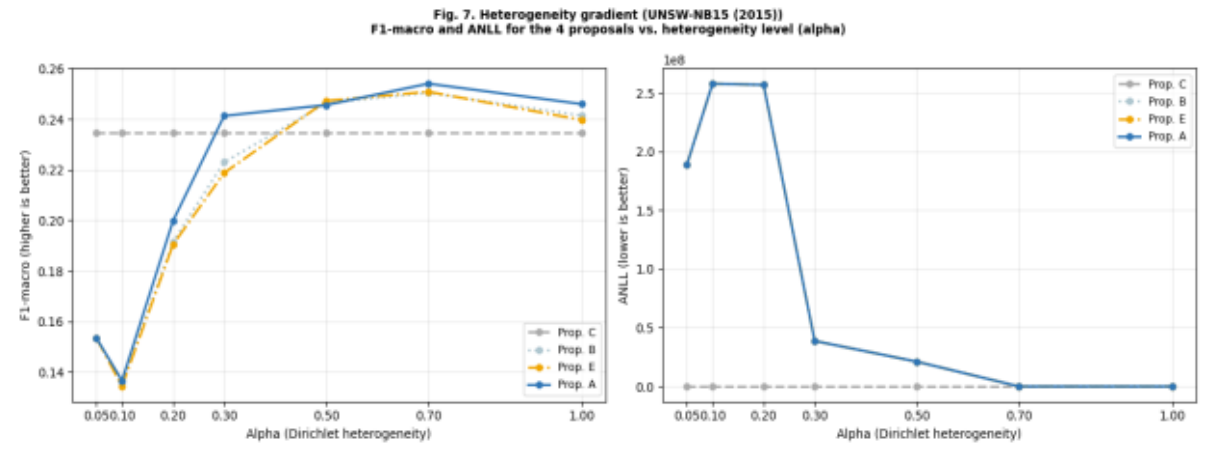

***Fig. 7.*** *Heterogeneity gradient, UNSW-NB15 (2015). The centralized ceiling (gray) outperforms federated proposals due to rare-class disappearance under Dirichlet partitioning.*

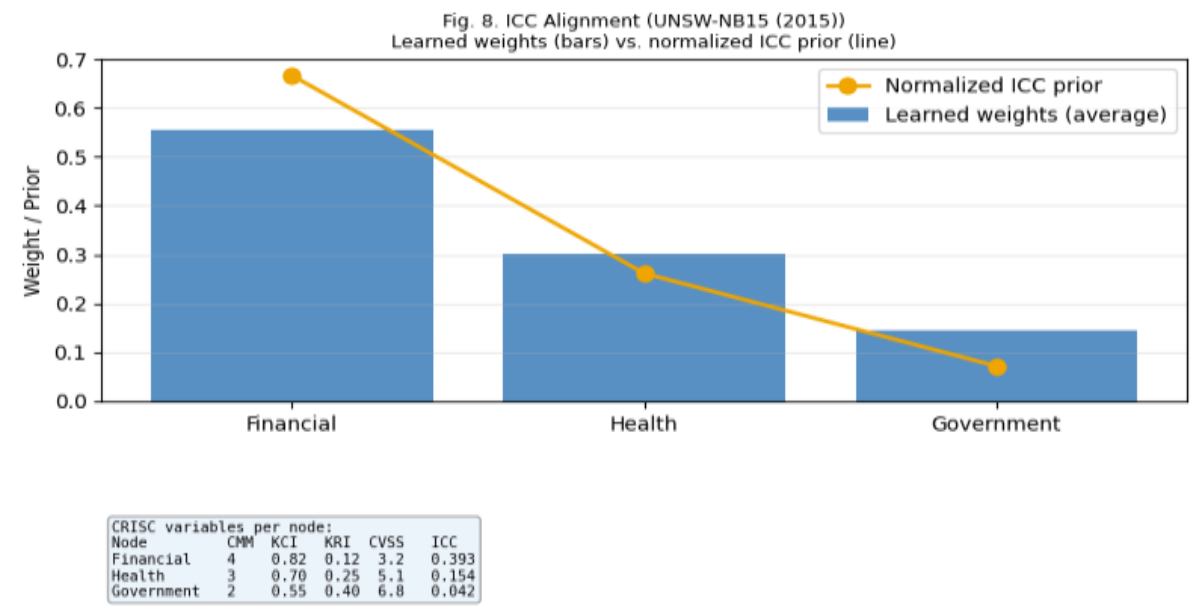

***Fig. 8.*** *ICC Alignment, UNSW-NB15 (2015). Clearest alignment with the full ICC hierarchy of any dataset: Financial > Health > Government.*

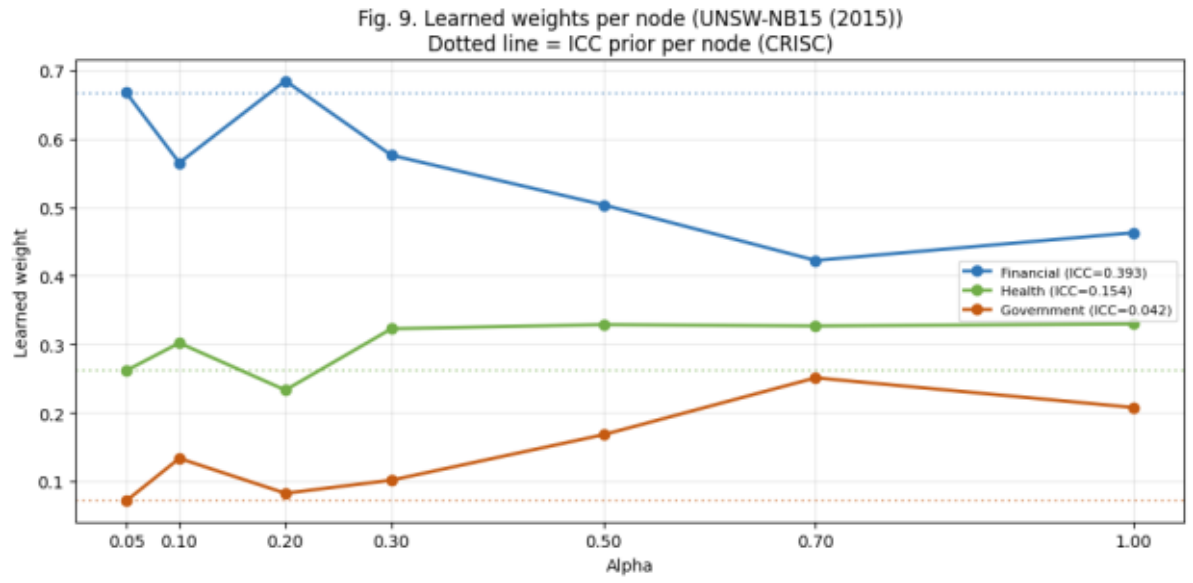

***Fig. 9.*** *Learned weights per node, UNSW-NB15 (2015). Financial node (blue) maintains dominant weight across all seven heterogeneity levels.*

### ***D.*** *ICC Alignment*

In every dataset and every heterogeneity level, the Financial node (ICC = 0.393) received a higher learned weight than the Government node (ICC = 0.042). No explicit ordering constraint was placed in the objective function. The optimizer arrived at this result by minimizing the composite loss, with the ICC prior as a regularization term. In all three cases, governance hierarchy and learned weights aligned.

**TABLE VII.** ICC Alignment: Learned Weights (avg. 5 reps x 7 levels)

| Dataset | w(Fin.) | w(Hlth.) | w(Gov.) | Alignment |
|---|---|---|---|---|
| NSL-KDD | 0.374 | 0.330 | 0.296 | Fin > Gov |
| CIC-IDS2017 | ~0.40 | ~0.40 | ~0.20 | Fin > Gov |
| UNSW-NB15 | 0.552 | 0.301 | 0.148 | Full hierarchy |
| Combined | 0.374 | 0.330 | 0.296 | Fin > Gov: 3/3 |

### ***E.*** *Combined Summary*

**TABLE VIII.** All Datasets: A vs. B

| Dataset | A (ICC) | B (FedAvg) | Delta | McNemar |
|---|---|---|---|---|
| NSL-KDD 2009 | 0.9135 | 0.9076 | +0.006 | 6/7 |
| CIC-IDS2017 2017 | 0.7556 | 0.6771 | +0.079 | 6/7 |
| UNSW-NB15 2015 | 0.2110 | 0.2060 | +0.005 | 1/7 |
| Combined avg. | 0.6049 | 0.5777 | +0.027 | 70/94 (74%) |

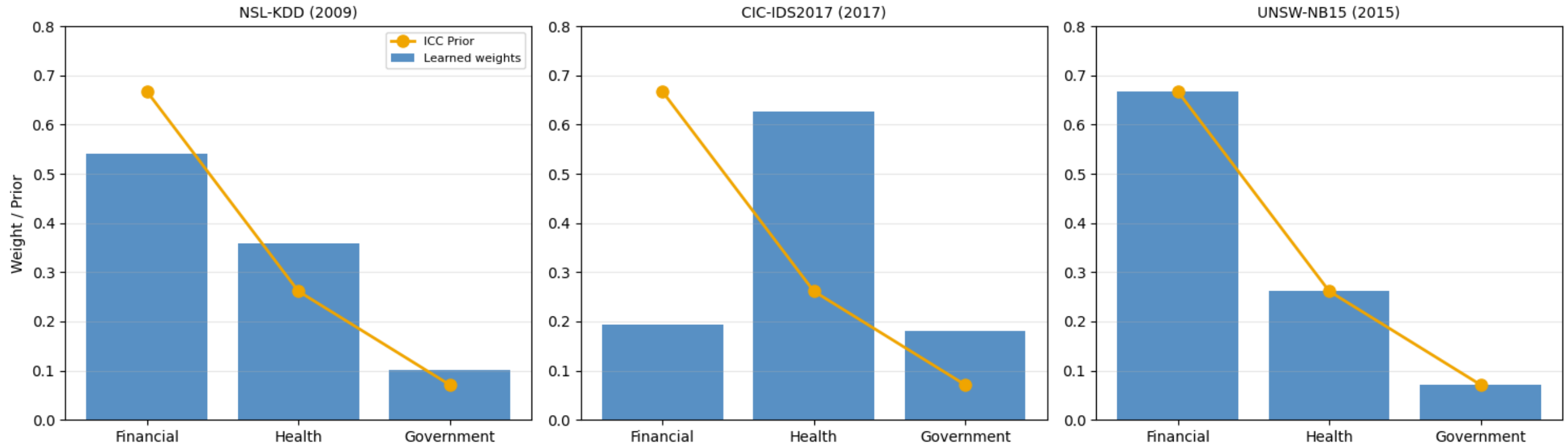

*Fig. 10.* *ICC Alignment cruzado: NSL-KDD (2009), CIC-IDS2017 (2017), UNSW-NB15 (2015). Bars: learned weights averaged over 5 reps and 7 alpha levels. Orange line: normalized ICC prior. In all three panels, the Financial node (left) consistently exceeds the Government node (right). The governance hierarchy learned by the optimizer matches the CRISC hierarchy in every case.*

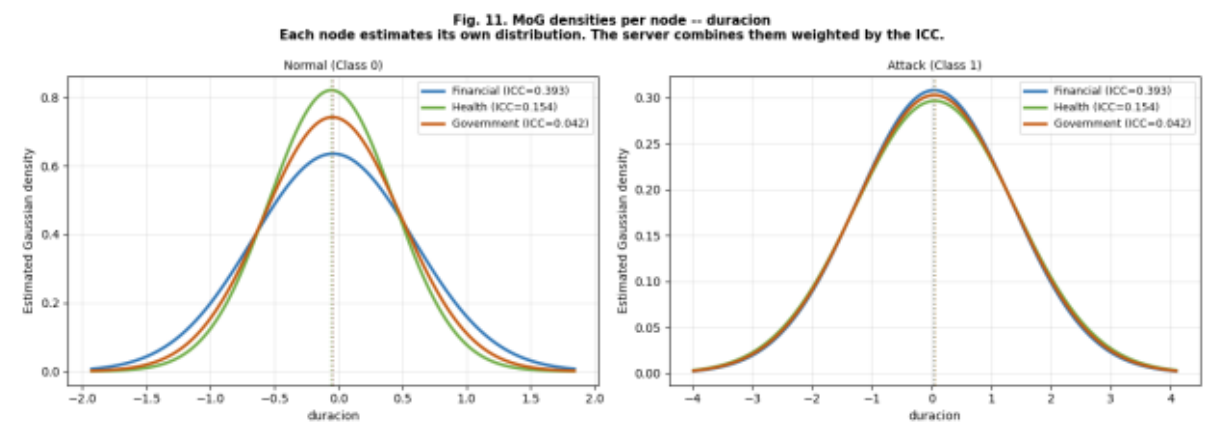

*Fig. 11.* *MoG densities per institutional node, NSL-KDD (2009), feature: duration. Left: Normal traffic. Right: Attack traffic. Each curve is one node's local Gaussian. The server weights these three distributions by the learned ICC weights at prediction time.*

## VI. DISCUSSION

### *A. Performance Consistency Across Configurations*

Proposal A does not produce statistically significant degradation relative to B in any of the 94 configurations tested. Confirmed improvement appears in 70 of them (74%). For a practitioner deciding whether to deploy this mechanism, the absence of degradation across three different datasets is probably the more relevant number.

From an operational standpoint, a federated model that reliably improves without ever reliably harming is deployable. The question auditors and CISOs actually ask is not how large the gain is; it is whether the new mechanism introduces risk. In all three environments tested, it does not.

### *B. Numerical Scaling in the Optimizer Objective*

Earlier versions of this program produced near-zero weight differences between proposals. The root cause was a scale mismatch inside the Nelder-Mead objective: ANLL values reached the millions while the ICC regularization term operated between 0 and 0.5. The optimizer was minimizing prediction loss with no effective governance signal. Applying log-softmax normalization to the ANLL before computing the objective brought both terms to the same numerical order, and the ICC prior became visible to the optimizer. The alignment pattern in Fig. 10 is a direct consequence of that correction.

### *C. UNSW-NB15: Structural Constraints Under Rare-Class Partitioning*

The low significance on UNSW-NB15 (1/7 levels) is not a failure of the ICC mechanism. It is a structural property of federated Naive Bayes with rare multiclass categories. When Dirichlet partitioning removes all Worms samples from a node's training set, GaussianNB cannot estimate mean and variance for that class in that node. The MoG server then combines two reliable distributions with one degenerate one. The ICC prior cannot fix a density estimation problem caused by absent data.

Federated generative models under severe Non-IID conditions face this structural ceiling, and addressing it is the clearest direction for future work.

### *D. Cross-Dataset Generalization*

Anyone who has deployed intrusion detection in production knows that a result on one dataset proves almost nothing. Network traffic from a university testbed in 2009 looks nothing like traffic from a Canadian cybersecurity institute in 2017, and nothing like traffic from a military cyber range in Australia in 2015. Different attack families, different equipment, and labeling conventions developed independently by different research groups. The fact that the same governance prior, the same optimizer, and the same parameters produce the same directional result in all three environments is the argument that the ICC mechanism captures something real.

## VII. CONCLUSIONS

Federated weight learning regularized by CRISC governance indicators outperforms size-proportional federated averaging in all three network environments evaluated. The combined F1-macro improvement is +0.0272 with statistical significance in 74% of the 94 configurations tested.

The central finding is the ICC Alignment pattern. Without any explicit ordering constraint, the optimizer assigns higher weight to the node with the highest governance maturity and lower weight to the node with the lowest, across three datasets collected between 2009 and 2017 by different research groups on different continents.

Two technical decisions made this possible. First, the hybrid local classifier separates nominal and numerical features, eliminating the distance bias that comes from encoding protocol names as integers. Second, log-softmax normalization of the ANLL inside the optimizer objective brings the prediction loss and the ICC regularization term to the same numerical scale. Without that normalization, the governance prior is invisible. With it, CRISC information reaches the optimization trajectory.

Future work will address three open points: constrained simplex projection for strict weight floor enforcement, adaptive regularization strength that responds to observed heterogeneity at each alpha level, and extension to federations with more than three nodes.

## ACKNOWLEDGMENT

The author thanks Prof. Ezequiel Lopez Rubio and Prof. Juan Miguel Ortiz de Lazcano Lobato, Department of Languages and Computer Science, University of Malaga, for their methodological direction throughout this research. The hybrid CategoricalNB plus GaussianNB architecture, the OOD slot correction [April 24, 2026], the real MoG server requirement [April 16, 2026], the seven-level heterogeneity gradient [April 20, 2026], and the McNemar and ANLL metrics were each introduced or validated through their specific recommendations during the period March to May 2026.

## REFERENCES

[1] H. B. McMahan et al., "Communication-efficient learning of deep networks from decentralized data," AISTATS 2017. arXiv:1602.05629

[2] ISACA, CRISC Review Manual, 8th ed. ISACA, 2025.

[3] N. Moustafa and J. Slay, "UNSW-NB15: a comprehensive data set for network intrusion detection systems," IEEE MilCIS, 2015. DOI: 10.1109/MilCIS.2015.7348942

[4] N. Moustafa and J. Slay, "The evaluation of Network Anomaly Detection Systems," Information Security Journal, 2016. DOI: 10.1080/19393555.2015.1125974

[5] I. Sharafaldin et al., "Toward generating a new intrusion detection dataset," ICISSP 2018. DOI: 10.5220/0006639801080116

[6] M. Tavallaee et al., "A detailed analysis of the KDD CUP 99 data set," IEEE CISDA, 2009. DOI: 10.1109/CISDA.2009.5137363

[7] T. Wisanwanichthan and M. Thammawichai, "Double-layered hybrid NID using Naive Bayes and SVM," IEEE Access, vol. 9, 2021. DOI: 10.1109/ACCESS.2021.3118573

[8] A. Lara-Gutierrez et al., "Framework for Drift Detection in AI-driven Anomaly Detection," Int. J. Inf. Security, vol. 24, 2025. DOI: 10.1007/s10207-025-01118-9

[9] K. A. ElDahshan et al., "Meta-heuristic optimization hierarchical IDS," Computers, vol. 11, 2022. DOI: 10.3390/computers11120170